\documentclass[%
 reprint,
superscriptaddress,
 amsmath,amssymb,
 aps,
twocolumn
12pt,margin=1.5in,
]{revtex4-1}
\usepackage{amsmath}

\usepackage{graphicx}
\usepackage{soul}
\usepackage{mhchem, textcomp}
\usepackage{graphicx}
\usepackage{dcolumn}
\usepackage{bm}
\usepackage{wrapfig}
\usepackage{graphicx, type1cm, lettrine}
\usepackage{float}
\usepackage{booktabs}
\usepackage{adjustbox}
\restylefloat{figure}
\usepackage{hyperref}
\usepackage{array}
\newcolumntype{P}[1]{>{\centering\arraybackslash}p{#1}}

\begin{document}

\preprint{APS/123-QED}

\title{Photoresponsivity enhancement in monolayer MoS$_2$ by rapid O$_2$:Ar plasma treatment}

\author{Jakub Jadwiszczak}
\affiliation{School of Physics, Trinity College Dublin, Dublin 2, Ireland}
\affiliation{Centre for Research on Adaptive Nanostructures and Nanodevices (CRANN) and Advanced Materials and Bioengineering Research (AMBER) Research Centers, Trinity College Dublin, Dublin 2, Ireland.}

\author{Gen Li}
\affiliation{School of Physics, Trinity College Dublin, Dublin 2, Ireland}
\affiliation{Centre for Research on Adaptive Nanostructures and Nanodevices (CRANN) and Advanced Materials and Bioengineering Research (AMBER) Research Centers, Trinity College Dublin, Dublin 2, Ireland.}

\author{Conor P. Cullen}
\affiliation{Centre for Research on Adaptive Nanostructures and Nanodevices (CRANN) and Advanced Materials and Bioengineering Research (AMBER) Research Centers, Trinity College Dublin, Dublin 2, Ireland.}
\affiliation{School of Chemistry, Trinity College Dublin, Dublin 2, Ireland}

\author{Jing Jing Wang}
\affiliation{Centre for Research on Adaptive Nanostructures and Nanodevices (CRANN) and Advanced Materials and Bioengineering Research (AMBER) Research Centers, Trinity College Dublin, Dublin 2, Ireland.}

\author{\\Pierce Maguire}
\affiliation{School of Physics, Trinity College Dublin, Dublin 2, Ireland}
\affiliation{Centre for Research on Adaptive Nanostructures and Nanodevices (CRANN) and Advanced Materials and Bioengineering Research (AMBER) Research Centers, Trinity College Dublin, Dublin 2, Ireland.}

\author{Georg S. Duesberg}
\affiliation{Centre for Research on Adaptive Nanostructures and Nanodevices (CRANN) and Advanced Materials and Bioengineering Research (AMBER) Research Centers, Trinity College Dublin, Dublin 2, Ireland.}
\affiliation{School of Chemistry, Trinity College Dublin, Dublin 2, Ireland}
\affiliation{Institute of Physics, EIT 2, Faculty of Electrical Engineering and Information Technology, Universit\"{a}t der Bundeswehr M\"{u}nchen, Werner-Heisenberg-Weg 39, 85577 Neubiberg,
Germany}

\author{James G. Lunney}
\affiliation{School of Physics, Trinity College Dublin, Dublin 2, Ireland}

\author{Hongzhou Zhang}
\email{hozhang@tcd.ie}
\affiliation{School of Physics, Trinity College Dublin, Dublin 2, Ireland}
\affiliation{Centre for Research on Adaptive Nanostructures and Nanodevices (CRANN) and Advanced Materials and Bioengineering Research (AMBER) Research Centers, Trinity College Dublin, Dublin 2, Ireland.}

\begin{abstract}

We report up to ten-fold enhancement of the photoresponsivity of monolayer MoS$_2$ by treatment with O$_2$:Ar (1:3) plasma. We characterize the surface of plasma-exposed MoS$_2$ by TEM, Raman and PL mapping and discuss the role of MoO$_x$ in improving the photocurrent generation in our devices. At the highest tested laser power of 0.1 mW, we find ten-fold enhancements to both the output current and carrier field-effect mobility under the illumination wavelength of 488 nm. We suggest that the improvement of electrical performance is due to the surface presence of MoO$_x$ resulting from the chemical conversion of MoS$_2$ by the oxygen-containing plasma. Our results highlight the beneficial role of plasma treatment as a fast and convenient way of improving the properties of synthetic 2D MoS$_2$ devices for future consideration in optoelectronics research.
\end{abstract}

\maketitle 
\clearpage

\lettrine[lines=2]{T}wo-dimensional layered transition metal dichalcogenides (TMDs) have attracted wide research interest due to their intriguing physical properties and potential applications. Molybdenum disulfide (MoS$_2$), a typical layered TMD, is a semiconductor with a direct bandgap of $\sim$ 1.8 eV in the single-layer limit \cite{Mak2010a}. This allows monolayer MoS$_2$ field-effect transistors (FETs) to achieve high ON/OFF ratios \cite{qiu2012electrical}  (10$^7$-10$^9$), making them attractive candidates for switching components in future electronics. Recently, optoelectronic devices fabricated from MoS$_2$ have received notable attention \cite{LopezSanchez2013, Chen2015, qin2016atomic,wang2018recent}. MoS$_2$ phototransistors are easy to fabricate, respond to a wide range of wavelengths \cite{LopezSanchez2013, Wang2015}, and exhibit fast DC photoresponses \cite{yore2017large,wang2015ultrafast}. In addition, their photoresponsivity can be tuned by various methods, such as back-gating \cite{Yin2012,lee2017wavelength}, encapsulation in HfO$_2$ \cite{Kufer2015}, strain engineering \cite{wang2017thermally}, layer decoupling \cite{yang2017sensitized} and evaporation of sub-stoichiometric molybdenum oxide overlayers \cite{Yoo2017}. Surface sensitization of monolayer MoS$_2$ FETs has also yielded significant enhancements of the measured photocurrent in the case of quantum dots \cite{kufer2015hybrid,kufer2016interface,gough2018dependence}, organic molecules \cite{yu2014dye,kang2015high,huang2016effects} and metal nanostructures \cite{Miao2015,jing2017ag}. However, these methods often involve additional preparation steps in order to fabricate the sensitizing species and deposit it on the MoS$_2$ device. Moreover, the surface-deposited dopants may not be robust to mechanical stressing or further material modification without losing their favorable properties.

Plasma functionalization, in turn, presents a fast and facile way to alter the crystal structure of on-chip layered materials such as MoS$_2$. It facilitates large-scale, multi-sample and rapid tuning of the optoelectronic performance of FETs based on layered semiconductors. In particular, oxygen-containing plasmas tend to form sub-stoichiometric molybdenum oxides on the surface of MoS$_2$ \cite{islam2014tuning,giannazzo2017ambipolar,ko2016stack}. These oxide centres can then act as dopants that alter the charge concentration in the modified MoS$_2$ transistor channel \cite{choudhary2016two,jadwiszczak2018oxide,jadwiszczak2018low}, and ultimately govern the electron conduction behavior of the newly-formed oxide/MoS$_2$ heterostructure \cite{khondaker2016bandgap}. In this work, we demonstrate the enhancement of the photoresponsivity of chemical vapour deposition (CVD)-grown monolayer MoS$_2$ by O$_2$:Ar (1:3) plasma treatment. The photoresponsivity is improved ten-fold in gated devices after 2 seconds of exposure to the plasma. At the same time, the field-effect mobility of the device under illumination improves by over one order of magnitude. We carry out transmission electron microscopy (TEM) imaging and spectroscopic mapping to characterize the sample after plasma exposure, and attribute the observed photoresponse to the suppressed charge recombination mediated by surface-bound molybdenum oxides.

MoS$_2$ samples were synthesized on SiO$_2$/Si substrates using the CVD method previously reported \cite{Maria2014}. The flake thickness was confirmed by optical microscopy and Raman spectroscopy. Standard  electron beam lithography was carried out to fabricate the FET devices using PMMA resist and development in MIBK:IPA (1:3) solution. This was followed by metallization with Ti(10 nm)/Au(40 nm) contacts and lift-off in acetone. Plasma treatment was carried out in a Fischione Instruments 1020 plasma cleaner for 2 seconds, utilizing O$_2$:Ar (1:3) gas at a chamber pressure of $\sim$ 5 mbar. The electrical testing was performed at room temperature in a two-probe configuration (Imina miBot) using a source meter unit (Agilent B2912A) in the ambient. The devices were back-gated through the heavily p-doped Si substrate underneath the 285 nm SiO$_2$ overlayer. A 488 nm laser was used for irradiation. Its power density was tuned at five different levels and controlled to ensure no power fluctuation throughout the experiment. The laser was directed through a condenser lens (20$\times$, NA = 0.4) and the spot size was $\approx$ 1.5 $\mu$m. TEM was carried out in a FEI Titan 80-300 system operated at 300 kV, at a chamber pressure of $4 \times 10^{-7}$ mbar. Monolayer samples were transferred onto copper TEM grids using the polymer stamp transfer method \citep{bie2011site}. Fabricated devices were imaged in a Zeiss Nanofab helium ion microscope at a beam energy of 25 keV. Raman and photoluminescence (PL) spectra were acquired using a WITec Alpha 300R system ($\lambda$ = 532 nm). Raman spectra were acquired using a spectral grating with 1800 lines/mm. PL spectra were collected using a 600 lines/mm grating. A low laser power ($<$ 100 $\mu$W) was used during mapping to minimize any laser-induced damage or heating of the sample.

Figure 1(a) is a false-color helium ion micrograph of a typical contacted device. The contacts in our devices were always deposited in a parallel geometry, as visible in the image. The transistor channel length was over 5 $\mu$m to confine the laser irradiation solely to the MoS$_2$ region. This was done to avoid any heating effects on the metal/semiconductor interface which are known to induce p-type doping in intrinsic n-type TMDs \cite{seo2018writing}. We collected the output and transfer characteristics of the device under 5 different illumination powers. As the laser spot avoided the Au electrodes during illumination, we assume that all of the measured photo-generated current originated in the MoS$_2$ semiconductor.

\begin{figure}[]
    \centering
\includegraphics[width=85mm]{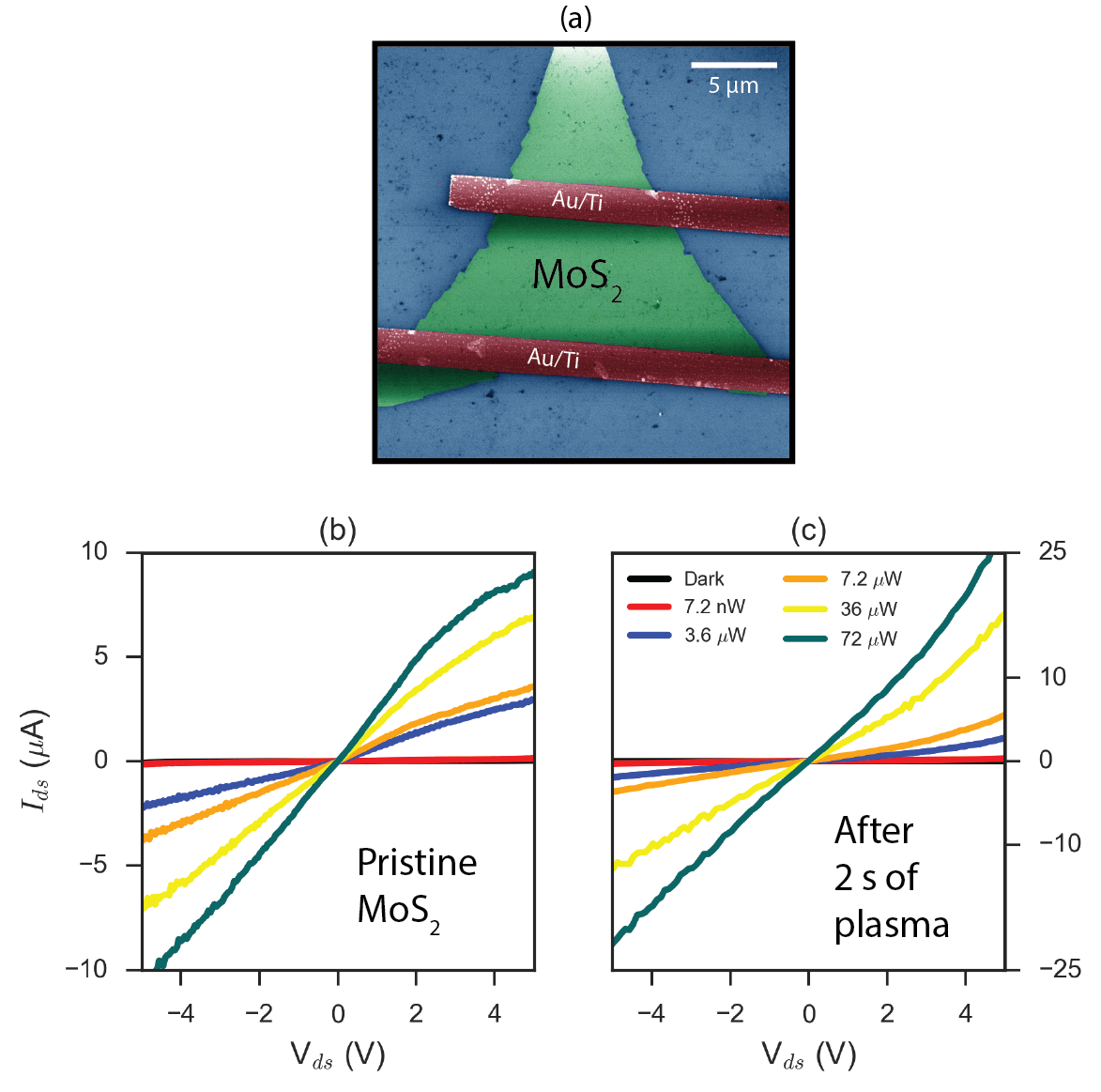}
\caption{{\footnotesize (a) False-color helium ion micrograph of a typical contacted monolayer MoS$_2$ phototransistor device. Blue area is the SiO$_2$ substrate. (b) Output curves of the untreated device, demonstrating a good ohmic contact between the material and the metal electrode. Output current, I$_{ds}$ increases with increasing laser power under illumination. (b) Post-plasma treatment IV curves show a similar trend with increasing laser power. The generated photocurrent at high laser powers has increased by up to 3 times at the same applied voltage after 2 seconds of plasma exposure. We note that no gate bias was applied. The color legend applies to both (b) and (c).}}
\end{figure}

Figure 1(b) shows the output characteristics of the device under laser illumination before any plasma treatment. Prior to any exposure to the plasma, the low-bias IV response of the MoS$_2$ FET shows a well-behaved linear increase with applied bias for both voltage polarities; indicating good ohmic contacts to the semiconductor. Upon successive irradiations with rising laser power, the photocurrent increases, which is typical for semiconducting monolayer MoS$_2$ devices \cite{Yin2012,LopezSanchez2013,zhang2013high,yore2017large,tang2017mos2,park2018near}. The output current reaches nearly 10 $\mu$A at $\pm$ 5 V at the highest tested laser power of 72 $\mu$W. Figure 1(c) tracks the IV curves after 2 seconds of exposure to the plasma. We see that the current increases to nearly 25 $\mu$A at the highest illumination power, compared with the untreated sample at the same applied drain-source voltage. This indicates that dopants introduced by the plasma treatment to the MoS$_2$ surface mediate an enhanced charge carrier photo-generation response in the device.

Figure 2(a) shows the transfer curves for the same sample before any plasma treatment. Our as-grown devices perform as standard n-type FETs with a field-effect mobility ($\mu$) of 0.13 cm$^2$ V$^{-1}$ s$^{-1}$ under no illumination, extracted in the linear region of the transfer curve and at V$_{ds}$ = 1 V. Upon successive laser irradiations we observe a photogating effect, whereby the threshold voltage of the transistor shifts to negative gate biases by more than 10 V due to increased electron doping. This has previously been observed in ultrathin TMD FETs and is attributed to the interaction of photo-generated carriers with charge traps in the transistor channel \cite{xie2017photodetectors,garcia2018photogating}. At the highest incident power, the FET channel is effectively still open at V$_{g}$ = - 60 V, where the output current stays firmly above 10$^{-7}$ A and leads to a large reduction in the ON/OFF ratio of our device.

\begin{figure}[]
  \centering
\includegraphics[width=85mm]{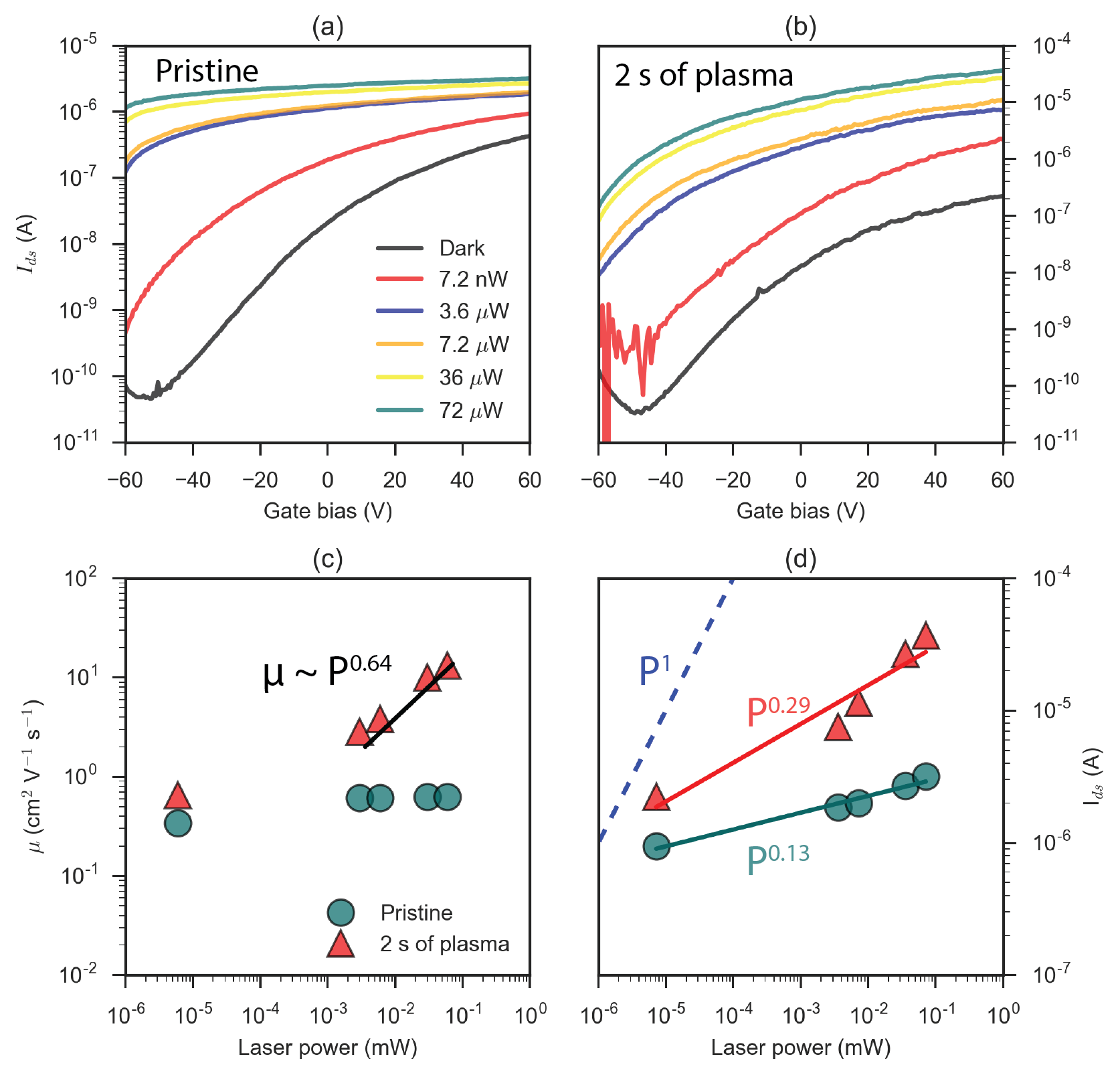}
\caption{{\footnotesize(a) Transfer characteristics of the untreated device, demonstrating standard n-type FET behaviour and increase of carriers in the channel at higher laser powers. (b) In the post-treatment gate curves, the level of current has increased by one order of magnitude at all illumination laser powers. The color legend applies to both (a) and (b). (c) Mobility comparison before and after plasma treatment as a function of laser power. The solid black line is a linear fit to the mobility scaling of the treated sample above 10$^{-3}$ mW. (d) Photocurrent comparison before and after plasma exposure as a function of 488 nm laser power. The power law fits to each data set are indicated on the plot. The blue dashed line shows an ideal P$^1$ response.}}
\end{figure}

Figure 2(b) presents the gate curves after plasma treatment. The observed level of output current in the dark transfer curve drops two-fold when evaluated at the gate bias, V$_{g}$ = 60 V. Meanwhile, the threshold voltage is seen to shift to more positive gate biases by $ \sim$ 5 V. This shift indicates oxygen-related p-type doping in the material, consistent with previous works on oxygen plasma-treated MoS$_2$ \cite{giannazzo2017ambipolar,guo2017observation,islam2014tuning,jadwiszczak2018low}. In addition, the MoS$_2$ now possesses a weak ambipolar response, indicating hole-branch conduction caused by the likely presence of plasma-created oxides \cite{chuang2014mos2, mcdonnell2014hole}. After 2 seconds of plasma treatment, the output current in the saturation region of the gate curve improves by one order of magnitude under all illumination powers (note scale on the y-axis). Figure 2(c) tracks the MoS$_2$ channel field-effect mobility before and after chemical reaction with the plasma. Even with no laser illumination, the mobility is seen to improve two-fold in the plasma-treated samples, which we have explored in previous work \cite{jadwiszczak2018oxide}.
After 2 seconds of exposure, the carrier mobility increases over ten-fold as the laser power is turned up. We find no clear relationship between the mobility and the laser power for the untreated sample. However, we obtain a good power law fit to the mobility scaling as $\mu \propto P^{0.64}$ above laser powers of 10$^{-3}$ mW. Similarly, in Fig. 2(d), the output current at V$_{g}$ = 60 V is seen to improve once the device is exposed to the plasma. In both the untreated and treated case, the dependence of the photocurrent on the laser power is sublinear, though the power law response is enhanced by plasma treatment from $\mu \propto P^{0.13}$ to $\mu \propto P^{0.29}$. The scaling exponent in this relationship depends on the charge trapping rate in the MoS$_2$ FET channel \cite{zhang2013high,massicotte2016picosecond}. Our results suggest that the presence of plasma-created oxides on the surface inhibits photo-generated pair recombination via defect sites. We extract both data sets at V$_g$ = 60 V where the FET is moving into depletion, i.e.: the majority carrier concentration in MoS$_2$ induced by gating begins to approach that of the photogenerated carrier density \cite{wu2013elucidating}. The slope of the fit to the photocurrent as a function of laser power serves as a measure of the photogating effect seen in the power-graded transfer curves in Figs. 2(a),(b). An increase in the slope after plasma treatment is a direct consequence of the additional charge present in the device.

\begin{figure}[]
    \centering
\includegraphics[width=85mm]{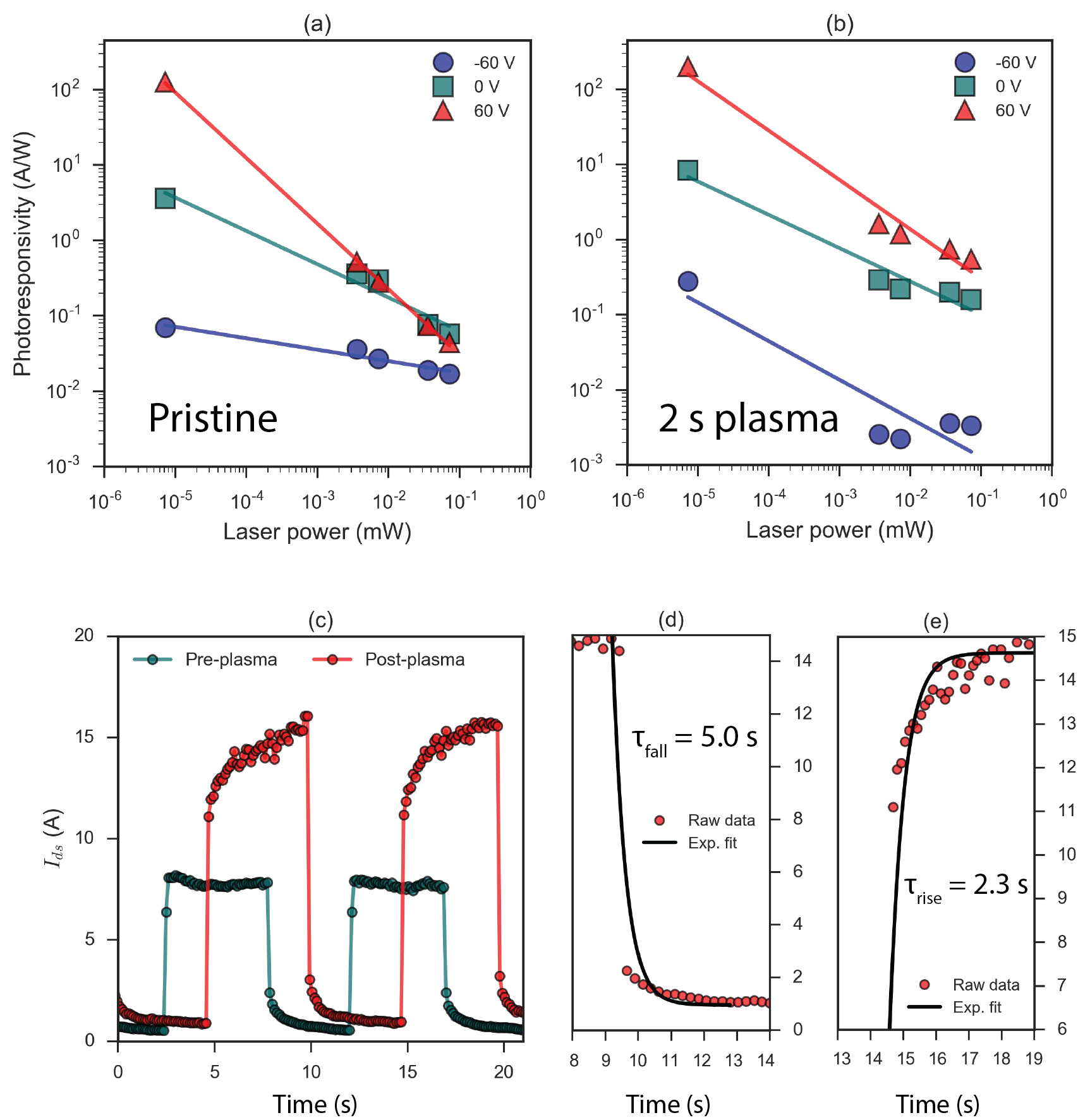}
\caption{{\footnotesize(a) $R_{ph}$ trends for the untreated MoS$_2$ sample. (b) Increased responsivity of the MoS$_2$ phototransistor after 2 seconds of O$_2$:Ar plasma treatment. (c) Comparison of temporal photocurrent response over laser irradiation cycles lasting 5 seconds. (d) Exponential fit of the fall component of the photoresponse for the treated device. (e) Exponential fit of the rise component from the next cycle.}}
\end{figure}

We plot the DC photoresponsivity, $R_{ph}$, at different gate biases as a function of irradiation power in Figure 3. $R_{ph}$ is the current generated in the device per unit of laser power and is a crucial parameter that quantifies the sensitivity of photodetectors \cite{xie2017photodetectors}. We obtain good linear fits of $R_{ph}$ as a function of power, $P$, across the whole gate bias range, before and after plasma treatment. The negative slope in the log-log plot indicates the saturation of trap states in the material with increasing incident optical power \cite{LopezSanchez2013, Yu2014,Wang2015}. In Fig. 3(a), we see the 0 V and 60 V gate bias trends exhibiting similar levels of $R_{ph}$, especially at higher laser powers. Upon plasma treatment, in Fig. 3(b), we observe an enhancement of $R_{ph}$ for all tested gate biases and a notable separation of the responsivity as a function of V$_g$. As V$_g$ is increased, the device becomes more responsive to laser illumination. The temporal response of the device pre- and post-plasma treatment is charted in Fig. 3(c). The photocurrent is seen to improve two-fold for the tested device when the laser irradiation is modulated through 5 s on/off cycles at a power of 36 $\mu$W and V$_{ds}$ = 5 V. The post-sensitization fall ($\tau_{fall}$) and rise ($\tau_{rise}$) times are extracted from single exponential fits in Figs. 3(d) and 3(e) respectively. The time-resolved photoresponses compare favorably with the evaporated MoO$_x$ overlayer report \cite{Yoo2017}, where our rise time at a much lower irradiation power is 35\% shorter.

\begin{center}
\begin{figure}[]
\includegraphics[width=85mm]{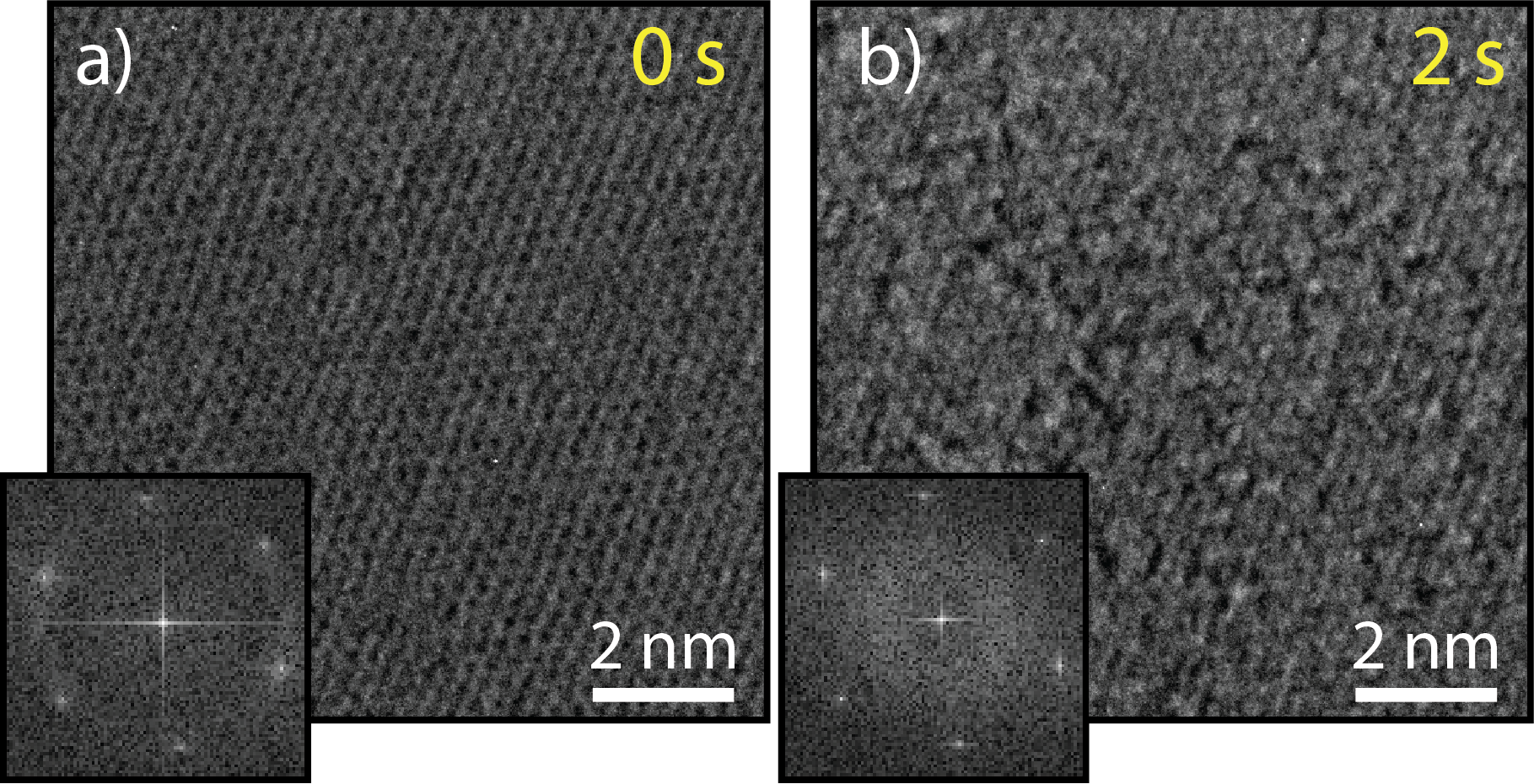}
\caption{{\footnotesize(a) TEM micrograph of a pristine monolayer of CVD MoS$_2$. (b) TEM image of the same MoS$_2$ flake after undergoing 2 seconds of plasma treatment. Insets show the FFTs for each image. }}
\end{figure}
\end{center}

Figures 5(a), (b) present TEM images of MoS$_2$ flakes before and after 2 s of plasma treatment. Corresponding Fast Fourier Transforms (FFTs) are included as insets. A large change in local contrast on some flake areas can be noted after 2 s of exposure to the plasma. It is evident from the TEM observation that after the plasma treatment, an amorphous oxide of molybdenum builds up on the surface as a consequence of a chemical reaction between the flake and the plasma-created species \cite{ko2016stack,nan2017improving,jadwiszczak2018oxide}. Spectroscopic mapping of the samples allows for a closer inspection of the chemical state of the MoS$_2$ surface pre- and post-plasma treatment. Figures 6(a), (b) show the spatially-resolved Raman maps of the material corresponding to the in-plane vibrational mode at 385 cm$^{-1}$. We notice a drastic drop in the intensity of the signal at this frequency, indicating a change in the MoS$_2$ lattice which alters the Raman-active modes in the sample. The flake-averaged spectra are presented in Fig. 6(c), demonstrating the quenching effect of plasma treatment on the monolayer MoS$_2$ Raman peaks.

From the spectral component fits (see Supplementary Table 1), the monolayer nature of the sample is confirmed with a wavenumber separation of 20.5 cm$^{-1}$ between the A$^{'}_{1}$ and E$^{'}$ peaks \cite{li2012bulk}. Upon plasma treatment, the intensity of both Raman modes is severely reduced after 2 seconds of exposure, while peak position also shifts and the full-width-at-half-maximum (FWHM) increases. Both the downshift of the E$^{'}$ peak and upshift of the A$^{'}_{1}$ peak are consistent with reports on molybdenum oxide formation on MoS$_2$, as is the asymmetric broadening of both peaks \cite{choudhary2016two,ko2016stack, jadwiszczak2018oxide}. PL maps of the neutral A exciton emission (1.84 eV) of the same flake are presented before and after 2 seconds of plasma treatment in Figs. 6(d), (e). Accompanying spectra averaged across the whole sample are shown in Fig. 6(f). We observe significant quenching of direct excitonic recombination in the sample after the plasma introduces the oxide species on the surface. The emission is also largely blue-shifted to higher energies by $\sim$ 0.1 eV. These observations are also in line with previous studies of oxidized MoS$_2$, where the emission intensity is reduced due to the presence of sub-stoichiometric oxides on the surface \cite{jadwiszczak2018oxide}.

\begin{figure}[]
    \centering
\includegraphics[width=85mm]{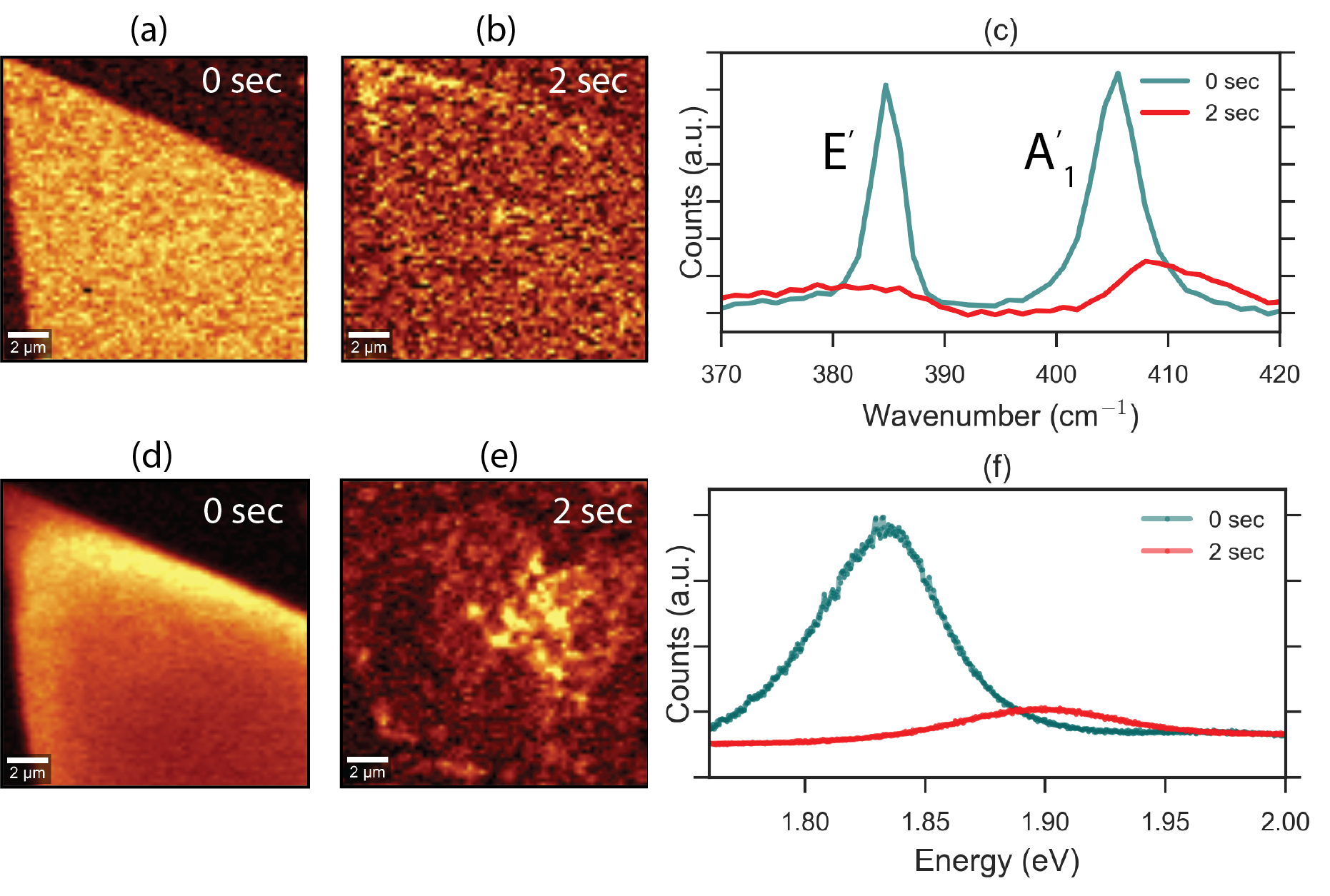}
\caption{{\footnotesize Raman maps of a monolayer flake before (a) and after (b) plasma treatment, filtered for the in-plane E$^{'}$ mode at 385 cm$^{-1}$. (c) Raman spectra averaged from the flake area. PL maps of the same flake before (d) and after (e) treatment tracking the direct A exciton emission. (e) Averaged PL spectra from the same flake area.}}
\end{figure}

We suggest that the observed photoresponsivity improvement results from carrier trapping at the MoS$_2$/MoO$_x$ interface \cite{Yoo2017}. The electron affinity and bandgap of monolayer MoS$_2$ are $\sim$ 4.3 eV and 1.8 eV respectively \cite{Liang2013,Mak2010a}. After the rapid plasma treatment, MoO$_x$ is generated on the device surface as demonstrated in the previous discussion. Oxides of molybdenum are commonly known as high work function materials (6.8 eV) with a bandgap of 3 eV \cite{Kang2014, Yoo2017}. In this device, plasma-generated oxides and unreacted MoS$_2$ will form an effective medium that spans the FET channel. As the Fermi level of MoS$_2$ is higher than that of MoO$_x$, significant band bending will occur at the interface \cite{Yoo2017}. The built-in electric field gradient will be directed from MoS$_2$ towards the oxide. Photo-generated holes will then become trapped at the material interface, inhibiting recombination and thereby enhancing the photocurrent with electrons as majority carriers. This is also supported by the observed photogating effect mediated by the electron-rich surface overlayer. The improved responsivity at higher back-gate fields is a direct consequence of Fermi level alignment which also facilitates easier photocarrier injection into the contacts \cite{Yin2012, Kufer2015, Chen2015} and primes the device for photon detection levels exceeding those of the pristine MoS$_2$.

In conclusion, we have demonstrated that the photoresponsivity of MoS$_2$ monolayer FETs can be enhanced ten-fold by the introduction of surface-bound molybdenum oxides. We confirm their presence via TEM, Raman and PL spectroscopy. The effect of the mobility and photoresponsivity enhancement depends on laser power and is more prominent at powers exceeding several $\mu$Watts. Our work provides insight into heterostructure physics in novel 2D optoelectronic nano-devices.

This research was supported in part by the Science Foundation Ireland (Grant nos: 11/PI/1105, 12/TIDA/I2433, 07/SK/I1220a and 08/CE/I1432) and the Irish Research Council (Grant nos: GOIPG/2014/972 and EPSPG/2011/239).


\bibliographystyle{apsrev}
\bibliography{my}

\end{document}